# Properties and Curie Temperature (130 K) of Heavily Mn-doped Quaternary Alloy Ferromagnetic Semiconductor (InGaMn)As Grown on InP

Shinobu Ohya[1,2], Hideo Kobayashi[1], and Masaaki Tanaka[1,2]
1. Department of Electronic Engineering, The University of Tokyo, 7-3-1 Hongo, Bunkyo-ku, Tokyo 113-8656, Japan
2. Science and Technology Corporation, 4-1-8 Honcho, Kawaguchi, Saitama 332-0012, Japan

We have studied magnetic properties of heavily Mn-doped $[(In_{0.44}Ga_{0.56})_{0.79}Mn_{0.21}]As$ thin films grown by low-temperature molecular-beam epitaxy (LT-MBE) on InP substrates. The (InGaMn)As with high Mn content (21%) was obtained by decreasing the growth temperature to 190°C. When the thickness of the $[(In_{0.44}Ga_{0.56})_{0.79}Mn_{0.21}]As$ layer is equal or thinner than 10 nm, the reflection high-energy electron diffraction (RHEED) pattern and transmission electron microscopy (TEM) show no MnAs clustering, indicating that a homogeneous single crystal with good quality was grown. In the magnetic circular dicroism (MCD) measurement, large MCD intensity and high Curie temperature of 130 K were observed.

III-V based diluted magnetic semiconductors (DMSs), such as (InMn)As[1] and (GaMn)As[2,3] are very attractive materials, because they have the properties of both semiconductors and magnetic materials and their spin-related functions can be used for novel "spintronic" devices. (These materials can give a new degree of freedom in the design of semiconductor materials and devices.) Many kinds of structures with magnetic semiconductors, such as magnetic quantum heterostructures[4,5] and magnetic tunnel junctions (MTJs)[6], have been intensively studied, and they were indicated to be very promising. Until recently, these studies mentioned above were limited to *ternary* alloys.

The *quaternary* alloy magnetic semiconductor (InGaMn)As has many potential advantages which cannot be realized by *ternary* alloy magnetic semiconductors. For example, the bandgap energy, easy magnetization axis and band structure can be controlled by changing the In content of (InGaMn)As. In particular, the bandgap is expected to be changed in a wide range, and can be tuned at the wavelength of 0.98, 1.3, or 1.55 μm for optical communication devices. Recently, we succeeded in growing $[(In_{0.53}Ga_{0.47})_{1-x}Mn_x]As$ on InP substrates by low-temperature molecular-beam epitaxy (LT-MBE) for the first time,[7] and have shown that 14% of Mn atoms can be incorporated in $[(In_{0.53}Ga_{0.47})_{1-x}Mn_x]As$ when grown at around 220°C.[8] High Curie temperature $T_C$ around 110 K of $[(In_{0.53}Ga_{0.47})_{0.87}Mn_{0.13}]As$ was also reported by T. Slupinsky et. al.[9] These results indicate that (InGaMn)As is also a very promising material.

In this work, we study the growth of heavily Mn-doped ferromagnetic $[(In_{0.44}Ga_{0.56})_{0.79}Mn_{0.21}]As$ containing the Mn concentration of 21%, which is much higher than the maximum Mn concentration (~7%) of ferromagnetic (GaMn)As, and is as high as that of (InMn)As. The magnetic properties and high $T_C$ of 130 K of this heavily Mn-doped (InGaMn)As film are reported below.

We grew $[(In_{0.44}Ga_{0.56})_{0.79}Mn_{0.21}]As$ films on InP (001) substrates by LT-MBE. The growth procedure was as follows. First, a 50-nm-thick $In_{0.44}Ga_{0.56}As$ buffer layer was grown on a semi-insulating InP (001) substrate at 500°C. This was followed by the LT-MBE growth of a 10-nm-thick $[(In_{0.44}Ga_{0.56})_{0.79}Mn_{0.21}]As$ film at very low temperature of 190°C. Lastly, we grew a 1-nm-thick $In_{0.44}Ga_{0.56}As$ cap layer. The In content was estimated by a x-ray diffraction peak of the InGaAs buffer layer. The Mn concentration was determined by extrapolating the Mn flux data obtained by x-ray measurements of (GaMn)As samples grown with various Mn cell's temperatures. After growth, the sample was annealed at low temperature of 220°C for 30 minutes in a $N_2$ atmosphere[10] in order to improve the magnetic properties of this (InGaMn)As sample.[11] When the thickness of the (InGaMn)As layer is equal or thinner than 10 nm, the reflection high-energy electron diffraction (RHEED) pattern during the growth was streaky with no MnAs clustering. In the growth of more than 10-nm-thick $[(In_{0.44}Ga_{0.56})_{0.79}Mn_{0.21}]As$, however, MnAs clustering was observed. All the results presented in this paper were measured on the sample with no MnAs clustering.

Figure 1 shows (a) a RHEED pattern after growth of a 10-nm-thick $[(In_{0.44}Ga_{0.56})_{0.79}Mn_{0.21}]As$ film when the electron beam azimuth is [110] and (b) a cross-sectional high-resolution transmission electron microscopy (TEM) image of this film taken with [110] projection. This TEM lattice image indicates that the (InGaMn)As layer has a zinc-blende type single crystal with no dislocations and with no visible MnAs clusters, despite its high Mn concentration of 21%.

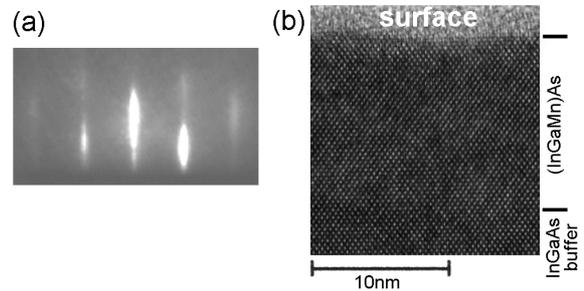

FIG. 1. (a) RHEED pattern after growth of an $[(In_{0.44}Ga_{0.56})_{0.79}Mn_{0.21}]As$ film with the electron beam azimuth of [110]. (b) A cross-sectional high-resolution transmission electron microscopy (TEM) image of the 10-nm-thick $[(In_{0.44}Ga_{0.56})_{0.79}Mn_{0.21}]As$ film with [110] projection. The (InGaMn)As layer has a zinc-blende type single crystal with no MnAs clustering and no dislocations.

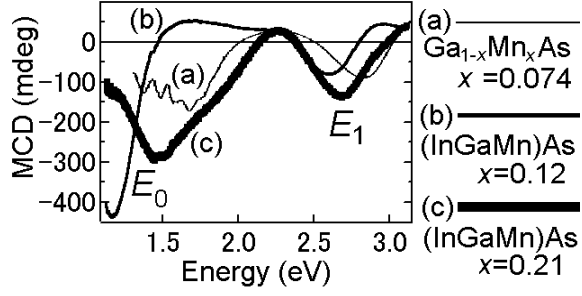

FIG. 2. MCD spectra of (a) an 1.4-μm-thick $Ga_{0.926}Mn_{0.074}As$ film, of (b) a 50-nm-thick $[(In_{0.53}Ga_{0.47})_{0.88}Mn_{0.12}]As$ film (Mn content $x=0.12$), and of (c) a 10-nm-thick $[(In_{0.44}Ga_{0.56})_{0.79}Mn_{0.21}]As$ film ($x=0.21$). (a) and (b) are measured as references. The spectrum of (a) was measured at 5 K, and those of (b) and (c) were measured at 10 K. All the spectra were measured with a magnetic field of 1 T applied perpendicular to the film. $E_0$ peaks of (b) and (c) were observed at 1.15 eV and 1.45 eV, respectively.

We carried out magnetic circular dicroism (MCD) measurements in a reflection setup. Here, the MCD intensity is expressed as

$$MCD[deg] = \frac{90}{\pi}\frac{R_+ - R_-}{R_+ + R_-} \propto \Delta E \frac{1}{R}\frac{dR}{dE}, \quad (1)$$

where $R_+$ and $R_-$ are optical reflectance for σ+ and σ- circularly polarized light, respectively, $\Delta E$ is Zeeman splitting energy which is proportional to the vertical component of magnetization, $R$ is the total reflectance, and $E$ is the electron energy. Since MCD is proportional to the vertical component of magnetization and to $dR/dE$, it provides the information both of the magnetization and the band structure. MCD measurements (spectra and magnetic filed dependence) are, therefore, very effective ways to investigate the properties of ferromagnetic semiconductors.

Figure 2 shows the MCD spectra of (a) an 1.4-μm-thick $Ga_{0.926}Mn_{0.074}As$ film, of (b) a 50-nm-thick $[(In_{0.53}Ga_{0.47})_{0.88}Mn_{0.12}]As$ film (Mn content $x=0.12$) and of (c) a 10-nm-thick $[(In_{0.44}Ga_{0.56})_{0.79}Mn_{0.21}]As$ film ($x=0.21$). (a) and (b) are measured as references. All the spectra were measured with a magnetic field of 1 T applied perpendicular to the film. Although the thickness of the $[(In_{0.44}Ga_{0.56})_{0.79}Mn_{0.21}]As$ film is very thin, large MCD intensity (maximum 300 mdeg.) was observed as shown in (c), and large MCD peaks of $E_0$ (1.45 eV) and $E_1$ (2.70 eV) were observed. This spectrum's feature is very similar to that of (a), meaning that the band structure of $[(In_{0.44}Ga_{0.56})_{0.79}Mn_{0.21}]As$ is of zinc-blende type like (GaMn)As. We can also see that the $E_0$ and $E_1$ peaks of (c) shift to higher energy from those of (b) observed at 1.15 eV and 2.59 eV, respectively. The bandgap energy of $In_{0.44}Ga_{0.56}As$ and of $In_{0.53}Ga_{0.47}As$ at 10 K are estimated to be 0.94 eV and 0.84 eV respectively, thus the energy difference between the bandgap energy of $In_{0.44}Ga_{0.56}As$ and the $E_0$ peak of $[(In_{0.44}Ga_{0.56})_{0.79}Mn_{0.21}]As$ ($x=0.21$) is 0.51 eV, and the energy difference between the bandgap energy of $In_{0.53}Ga_{0.47}As$ and the $E_0$ peak of $[(In_{0.53}Ga_{0.47})_{0.88}Mn_{0.12}]As$ ($x=0.12$) is 0.31 eV. This result implies that the blue shift of the MCD peak tends to become larger with increasing the Mn concentration. A similar peak's blue shift was observed also in (GaMn)As. One possible reason for this blue shift is the Moss-Burstein shift induced by the increase of the Fermi energy, which is higher than the valence band (in terms of hole energy), induced by the increase of the hole density with the increase of the Mn concentration $x$. In the case of (GaMn)As, the blue shift with increasing $x$ is smaller (around 0.1 eV), so the energy shift can be explained by the Moss-Burstein shift. In the case of this (InGaMn)As sample, however, the blue shift is so large that it is difficult to explain this result solely with the Moss-Burstein shift. Therefore the peak shift may be mainly due to the bandgap enlargement of (InGaMn)As with increasing the Mn concentration, which has been very difficult to be observed in III-V based DMS.

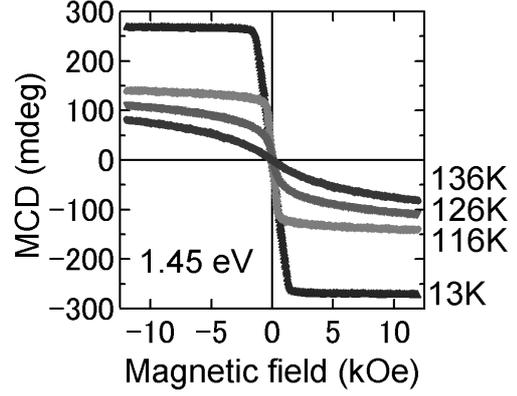

FIG. 3. Magnetic field dependence of MCD of an $[(In_{0.44}Ga_{0.56})_{0.79}Mn_{0.21}]As$ film at 13 K, 116 K, 126 K and 136 K measured at 1.45 eV. The magnetic field was applied perpendicular to the film.

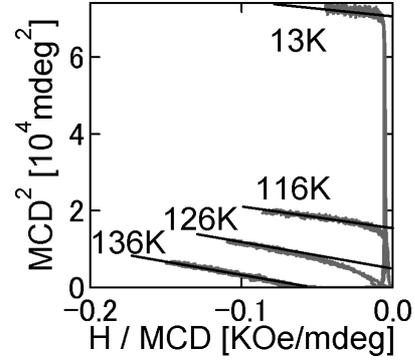

FIG. 4. The Arrott plot of an $[(In_{0.44}Ga_{0.56})_{0.79} Mn_{0.21}]As$ film from the MCD measurements. The estimated $T_C$ was 130 K.

We also measured the magnetic field dependence of the MCD intensity of the $[(In_{0.44}Ga_{0.56})_{0.79}Mn_{0.21}]As$ film at 1.45 eV at various temperatures with a magnetic field applied perpendicular to the film plane. This result is shown in Fig. 3. The ferromagnetic order was clearly observed even at 126 K. In order to estimate the Curie temperature ($T_C$) of this sample, we carried out the Arrott plot (MCD$^2$ vs. H/MCD), as shown in Fig. 4. By this plot, $T_C$ was estimated to be 130 K, which is quite high for magnetic alloy semiconductors. This $T_C$ is the higher than the values of (InGaMn)As$^{7-9}$ (~110 K) and of (InMn)As$^{12}$ (50 K) reported thus far.

By the measurement of temperature dependence of magnetic moment with a dc superconducting quantum interface device (SQUID) of the [(In$_{0.44}$Ga$_{0.56}$)$_{0.79}$Mn$_{0.21}$]As film, the Mn atoms contributing to the magnetic moment is estimated to be 46 % of the total number of the doped Mn atoms, when assuming that the spin of Mn is 5/2. This result is consistent with the experimentally obtained relationship between $T_C$ and the effective Mn concentration contributing to the magnetic moment for (InGaMn)As shown by T. Slupinski, et. al.$^9$

We also measured the anomalous Hall effect of an [(In$_{0.44}$Ga$_{0.56}$)$_{0.79}$Mn$_{0.21}$]As sample. (This sample is different from the one mentioned above, grown with slightly different conditions.) Clear ferromagnetic order was also observed. Figure 5 shows the annealing temperature dependence of $T_C$ estimated by the Arrott plot of the anomalous Hall effect of this sample. By the low temperature annealing, a significant improvement of $T_C$ from 75 K (as grown) to 125 K (annealed at 250˚C for 30 minutes) was observed. The highest $T_C$ (125 K) observed here was almost the same as that observed in MCD measurements. This large improvement of $T_C$ indicates that the low temperature annealing is also very effective for heavily Mn doped (InGaMn)As.

In conclusion, we have successfully grown [(In$_{0.44}$Ga$_{0.56}$)$_{0.79}$Mn$_{0.21}$]As thin films with high Mn content (21%) on an InP (001) substrate, which was enabled by decreasing the growth temperature to 190˚C. When the thickness of the [(In$_{0.44}$Ga$_{0.56}$)$_{0.79}$Mn$_{0.21}$]As layer is equal or thinner than 10 nm, the RHEED pattern during the growth remained streaky with no MnAs clustering, indicating that a homogeneous single crystal with good quality was obtained. No MnAs clusters were observed in TEM images. In the MCD measurement, the energy of the $E_0$ peak was 0.51 eV higher than the bandgap energy of In$_{0.44}$Ga$_{0.56}$As. This may be due to the blue shift of the bandgap energy by incorporating the Mn atoms into InGaAs. By the magnetic field dependence of MCD, very high $T_C$ of 130 K was estimated. This is the highest among the $T_C$ values reported in (InGaMn)As and (InMn)As thus far. We also revealed that the low temperature annealing is also very effective to increase $T_C$ of heavily Mn-doped (InGaMn)As.

Acknowledgement
   This work was supported by the PRESTO program of JST, Toray Science Foundation, and Research Projects of Grant-in-Aid for Scientific Research from MEXT.

References

[1] H. Munekata, H. Ohno, S. von Molnar, A. Segmuller, L. L. Chang, and L. Esaki, Phys. Rev. Lett. **63**, 1849 (1989); H. Ohno, H. Munekata, S. von Molnar, and L. L. Chang, J. Appl. Phys. **69**, 6103 (1991).

[2] H. Ohno, J. Magn. & Magn. Mater. **200**, 110 (1999); H. Ohno, A. Shen, F. Matsukura, A. Oiwa, A. Endo, S. Katsumoto, and Y. Iye, Appl. Phys. Lett. **69**, 363 (1996).

[3] M. Tanaka, J. Vac. Sci. & Technol. B**16**, 2267 (1998); M. Tanaka, T. Hayashi, T. Nishinaga, H. Shimada, H. Tsuchiya, and Y. Otuka, J. Cryst. Growth **175/176**, 1063 (1997).

[4] M. Tanaka, H. Shimizu, T. Hayashi, H. Shimada, and K. Ando, J. Vac. Sci. & Technol. A**18**, 1247 (2000).

[5] T. Hayashi, M. Tanaka, and A. Asamitsu, J. Appl. Phys. **87**, 4673 (2000).

[6] M. Tanaka and Y. Higo, Phys. Rev. Lett. **87**, 26602 (2001); Y. Higo, H. Shimizu and M. Tanaka, J. Appl. Phys. **89**, 6745 (2001).

[7] S. Ohya, Y. Higo, H. Shimizu, and M. Tanaka, Ext. Abstr. of the 48th Spring Meeting The Japan Society of Applied Physics and Related Societies, March 2001; S. Ohya, Y. Higo, H. Shimizu, J. M. Sun, and M. Tanaka, Jpn. J. Appl. Phys. **41**, L24 - L27 (2002); S. Ohya, Y. Higo, H. Shimizu, J. M. Sun, and M. Tanaka, cond-mat / 0111163 (2001).

[8] S. Ohya, H. Yamaguchi, and M. Tanaka, to be published in J. Superconductivity.

[9] T. Slupinski, H. Munekata, and A. Oiwa, Appl. Phys. Lett. **80**, 1592 (2002).

[10] T. Hayashi, Y. Hashimoto, S. Katsumoto, and Y. Iye, Appl. Phys. Lett. **78**, 1691 (2001).

[11] We revealed that the low temperature annealing is also very effective to improve the magnetic properties of (InGaMn)As. We presented the effect in S. Ohya, H. Yamaguchi, and M. Tanaka, Twelfth International MBE Conference MA1.5, San Francisco, Sept.15-20, 2002; S. Ohya and M. Tanaka, the 63rd Autumn Meeting, The Japan Society of Applied Physics and Related Societies 26p-ZA-15, Sept. 2002.

[12] T. Slupinski, A. Oiwa, S. Yanagi, and H. Munekata, J. Cryst. Growth **237-239**, 1326 (2002)

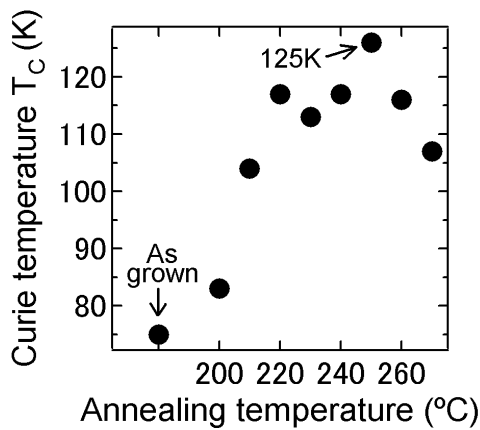

FIG. 5. Annealing temperature dependence of $T_C$. These $T_C$ values were estimated by the Arrott plot of the anomalous Hall effect of [(In$_{0.44}$Ga$_{0.56}$)$_{0.79}$Mn$_{0.21}$]As films.